# Interatomic correlations moments of atoms in the two-dimensional hexagonal lattice by using Morse and Lenard–Jones potentials

Clóves G. Rodrigues

*Escola de Ciências Exatas e da Computação, Pontifícia Universidade Católica de Goiás, Caixa Postal 86, 74605-010 Goiânia, Goiás, Brazil*

**ABSTRACT**

In this work we investigate the interatomic correlation moments in two-dimensional model of a weakly anharmonic crystal (i.e., not very high temperatures) with hexagonal lattice, using the Correlative Method of Unsymmetrized Self-Consistent Field (CUSF). The numerical results are obtained (and compared) by using the Morse and Lenard–Jones potentials.

## 1. Introduction

One of the most important characteristics of the lattice dynamics is the *quadratic correlation moments* between atomic displacements from lattice points [1,2]. In the harmonic approximation they have been calculated using the dynamical theory of crystal lattice [1]. It is well-known that anharmonic effects are significant at high temperatures [3].

Based on the Correlative Method of Unsymmetrized Self-Consistent Field (CUSF, for short) [3–7] general formulae were derived for quadratic correlation moments between atomic displacements from lattice points in crystals taking into account anharmonic terms up to the fourth order [8–10], and have used them for calculations of these characteristics in the linear chain [9,11], two-dimensional [10] and three-dimensional [12–17] models.

In this work we investigate the interatomic correlation moments in two-dimensional model of a weakly anharmonic crystal (i.e., not very high temperatures) with hexagonal lattice, using the Correlative Method of Unsymmetrized Self-Consistent Field (CUSF). The numerical results are obtained (and compared) by using the Morse and Lenard–Jones potentials.

## 2. General relations

In CUSF, the interatomic correlations moments between two atoms $i$ and $j$ in a crystal, is written as

$$C_{aa}(ij) = \overline{q_{ia} q_{ja}} \qquad (1)$$

with

E-mail address: cloves@pucgoias.edu.br

$$\mathbf{q}_i = \mathbf{r}_i - \hat{A}\mathbf{n}_i \qquad (2)$$

where $\hat{A}$ is the lattice matrix, $\mathbf{n}_i$ are the integer-components vectors and $a$ denotes the Cartesian components of atomic displacements. We consider a crystal with a pairwise central forces

$$U(\mathbf{r}_1, \mathbf{r}_2, \ldots, \mathbf{r}_N) = \frac{1}{2}\sum_{i\neq j}\Phi\big(|\mathbf{r}_i - \mathbf{r}_j|\big). \qquad (3)$$

In this case, taking into account anharmonic terms up to the fourth order we have of the interatomic correlations moments [9]

$$
\begin{aligned}
C_{ab}(ij) =\ & \frac{1}{\Theta}\Phi_{\alpha\beta}(ij)\overline{a_i\alpha_i}^0\overline{b_j\beta_j}^0 + \frac{1}{6\Theta}\Phi_{\alpha\beta\gamma\delta}(ij)\left(\overline{a_i\alpha_i\gamma_i\delta_i}^0\overline{b_j\beta_j}^0 - \overline{a_i\alpha_i}^0\overline{b_j\beta_j\gamma_j\delta_j}^0\right) \\
& + \frac{1}{\Theta^2}\sum_k \Phi_{\alpha\gamma}(ik)\Phi_{\beta\delta}(jk)\overline{a_i\alpha_i}^0\overline{b_j\beta_j}^0\overline{\gamma_k\delta_k}^0 \\
& - \frac{1}{4\Theta}\Phi_{\alpha\beta\gamma}(ij)\Phi_{\delta\epsilon\xi}(ij)\overline{a_i\alpha_i\gamma_i\delta_i}^0\overline{b_j\beta_j\epsilon_j\xi_j}^0 \\
& + \frac{1}{4\Theta^2}\sum_k \Phi_{\alpha\beta\gamma}(ik)\Phi_{\delta\epsilon\xi}(jk)\overline{a_i\alpha_i}^0\overline{b_j\delta_j}^0 \times \left(\overline{\beta_k\gamma_k\epsilon_k\xi_k}^0 - \overline{\beta_k\gamma_k}^0\overline{\epsilon_k\xi_k}^0\right) \\
& + \frac{1}{4\Theta^2}\Phi_{\alpha\beta\gamma}(ij)\sum_k \left(\Phi_{\delta\epsilon\xi}(jk)\overline{a_i\alpha_i}^0\overline{b_j\beta_j\gamma_j\delta_j}^0 - \Phi_{\delta\epsilon\xi}(ik)\overline{b_j\beta_j}^0\overline{a_i\alpha_i\gamma_i\delta_i}^0\right) \\
& - \frac{1}{4\Theta^2}\Phi_{\alpha\beta}(ij)\Phi_{\gamma\delta\epsilon\xi}(ij)\overline{a_i\alpha_i\gamma_i\delta_i}^0\overline{b_j\beta_j\epsilon_j\xi_j}^0 \\
& + \frac{1}{6\Theta^2}\sum_k \left(\Phi_{\alpha\gamma}(ik)\Phi_{\beta\delta\epsilon\xi}(jk) + \Phi_{\beta\gamma}(jk)\Phi_{\alpha\delta\epsilon\xi}(ik)\right)\overline{a_i\alpha_i}^0\overline{b_j\beta_j}^0\overline{\gamma_k\delta_k\epsilon_k\xi_k}^0 \\
& + \frac{1}{6\Theta^2}\sum_k \Big(\Phi_{\alpha\gamma}(ik)\Phi_{\beta\delta\epsilon\xi}(jk)\overline{a_i\alpha_i}^0\overline{b_j\beta_j\delta_j\epsilon_j}^0 \\
& + \Phi_{\beta\gamma}(jk)\Phi_{\alpha\delta\epsilon\xi}(ik)\overline{a_i\alpha_i\delta_i\epsilon_i}^0\overline{b_j\beta_j}^0\Big)\overline{\gamma_k\xi_k}^0, \qquad (4)
\end{aligned}
$$

where $\Theta = k_B T$ is the absolute temperature in energy units (here $k_B$ is as usual Boltzmann constant) and

$$\Phi_{\alpha\beta\ldots}(ij) = \left[\frac{\partial\cdots\Phi|\mathbf{r}|}{\partial x_\alpha \partial x_\beta \ldots}\right]_{\mathbf{r}=\hat{A}(\mathbf{n}_i-\mathbf{n}_j)}, \tag{5}$$

are the derivatives of the interatomic potential, and all Greek indices are dummy. For the sake of brevity, in the right-hand side, we write $a_i, \alpha_i, \ldots$ instead of $q_{ai}, q_{\alpha i}, \ldots$, and use the notation $\overline{a_i \alpha_i \ldots}^0$ for the moments in the zeroth-order approximation of the CUSF $\overline{q_{ia} q_{i\alpha} \ldots}^0$. Generally speaking, the summation in Eq. (4) extends over all $k$ except for $k=j$, and in the case of short-range forces one can restrict the summation over the nearest neighbors of the atom $i$ and $j$. These formulae are valid for any Bravais lattice of arbitrary dimensionality $n$. A specific lattice symmetry enables one to simplify them. For a perfect strongly anharmonic crystal of a high symmetry, the moments in the right side of Eq. (4) are expressed in terms of the solution $\beta_n(x)$ of the transcendental equation [18,19]

$$\beta_n(x) = nx \frac{D_{-(n/2+1)}\left[x+(n+2)\beta_n/2nx\right]}{D_{-n/2}\left[x+(n+2)\beta_n/2nx\right]}, \tag{6}$$

in which $D_\nu(z)$ are the parabolic cylinder functions and $x$ is a dimensionless combination of the temperature and the second- and fourth-order force coefficients.

In Eq. (4), the first two terms are of the first order and other terms are of the second order. One can see from Eq. (4) that the first order of CUSF enables one to calculate the correlations between nearest neighbors while the second order gives those between more distant atoms.

## 3. Two-dimensional hexagonal lattice

Low-dimensional models play an important part in developing theoretical methods. The two-dimensional hexagonal lattice represented in Fig. 1 is the simplest model of a closed-packed crystal. For it, Eq. (4) gives correlations between nearest, second and third neighbors. For any pair of atoms we shall use the coordinate system whose $X$-axis runs through the corresponding lattice points. In case of the second neighbors, its axes are denoted as $X'$, $Y'$ in Fig. 1.

We consider the case of a weak anharmonicity, i.e., not very high temperatures. Then, for a crystal with nearest-neighbor interactions and hexagonal lattice, we obtain

$$C_{xx}(1) = \frac{11\Theta}{108f}\left[1 - \frac{20\Theta}{33f^2}\left(h - \frac{31g^2}{80f} + \frac{g}{8r_0} - \frac{29f}{80r_0^2}\right)\right], \tag{7}$$

$$C_{xx}(2) = \frac{\Theta}{36f}\left[1 + \frac{\Theta}{3f^2}\left(\frac{3g^2}{4f} - h - \frac{5g}{2r_0} - \frac{29f}{12r_0^2}\right)\right], \tag{8}$$

$$C_{xx}(3) = \frac{\Theta}{27f}\left[1 - \frac{\Theta}{3f^2}\left(h - \frac{g^2}{2f} - \frac{3f}{2r_0^2}\right)\right], \tag{9}$$

where $r_0$ is the minimum point of the interaction potential and

$$f = \Phi^{(II)}(r_0), \quad g = \Phi^{(III)}(r_0), \quad h = \Phi^{(IV)}(r_0), \tag{10}$$

are the second, third and fourth derivatives in the point $r_0$ (of course, $e = \Phi^{(I)}(r_0) = 0$).

## 4. Interatomic potentials

For the numerical results we used the Morse and Lenard–Jones interatomic potentials. The Morse interatomic potential is

$$\Phi(r) = \varepsilon\left[e^{-2\rho(r-r_0)} - 2e^{-\rho(r-r_0)}\right], \tag{11}$$

where $\rho = 6/r_0$, $\varepsilon$ being the depth of the potential and $r_0$ is the minimum point of the interatomic potential.

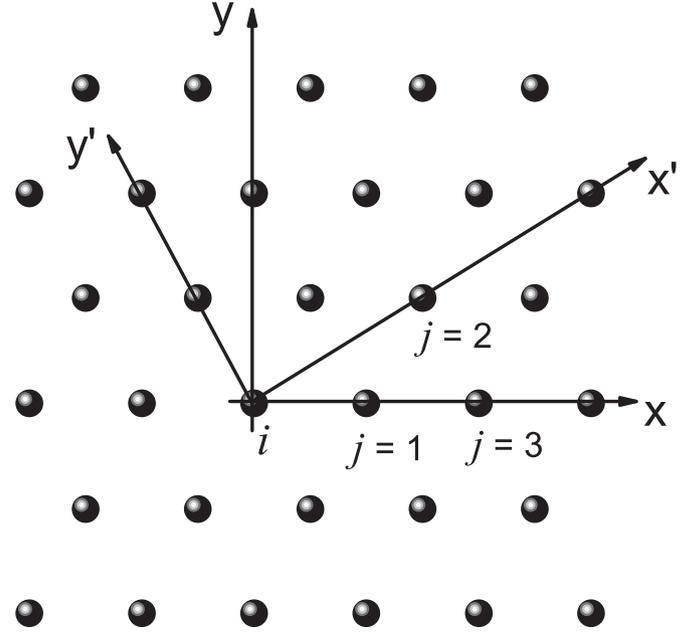

**Fig. 1.** A fragment of hexagonal lattice.

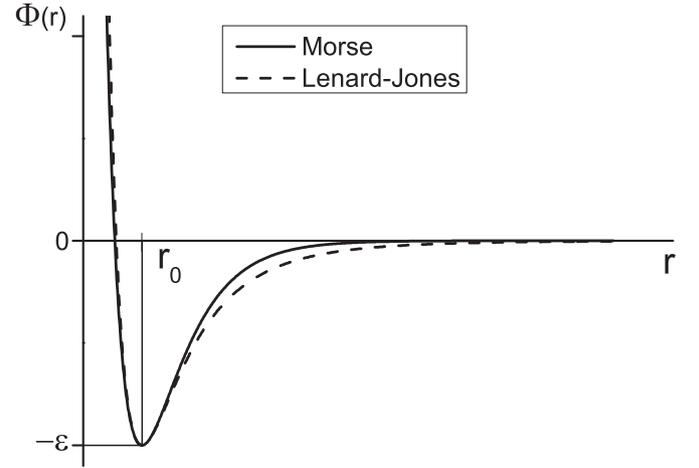

**Fig. 2.** Morse and Lenard–Jones potentials.

**Table 1**
Derivatives of the interatomic potentials.

| Derivatives | Morse | Lenard–Jones |
|---|---|---|
| $e$ | 0 | 0 |
| $f$ | $72\varepsilon/r_0^2$ | $72\varepsilon/r_0^2$ |
| $g$ | $-1296\varepsilon/r_0^3$ | $-1512\varepsilon/r_0^3$ |
| $h$ | $18144\varepsilon/r_0^4$ | $26712\varepsilon/r_0^4$ |

The Lennard–Jones potential is

$$\Phi(r) = 4\varepsilon\left[\left(\frac{\sigma}{r}\right)^{12} - \left(\frac{\sigma}{r}\right)^6\right], \tag{12}$$

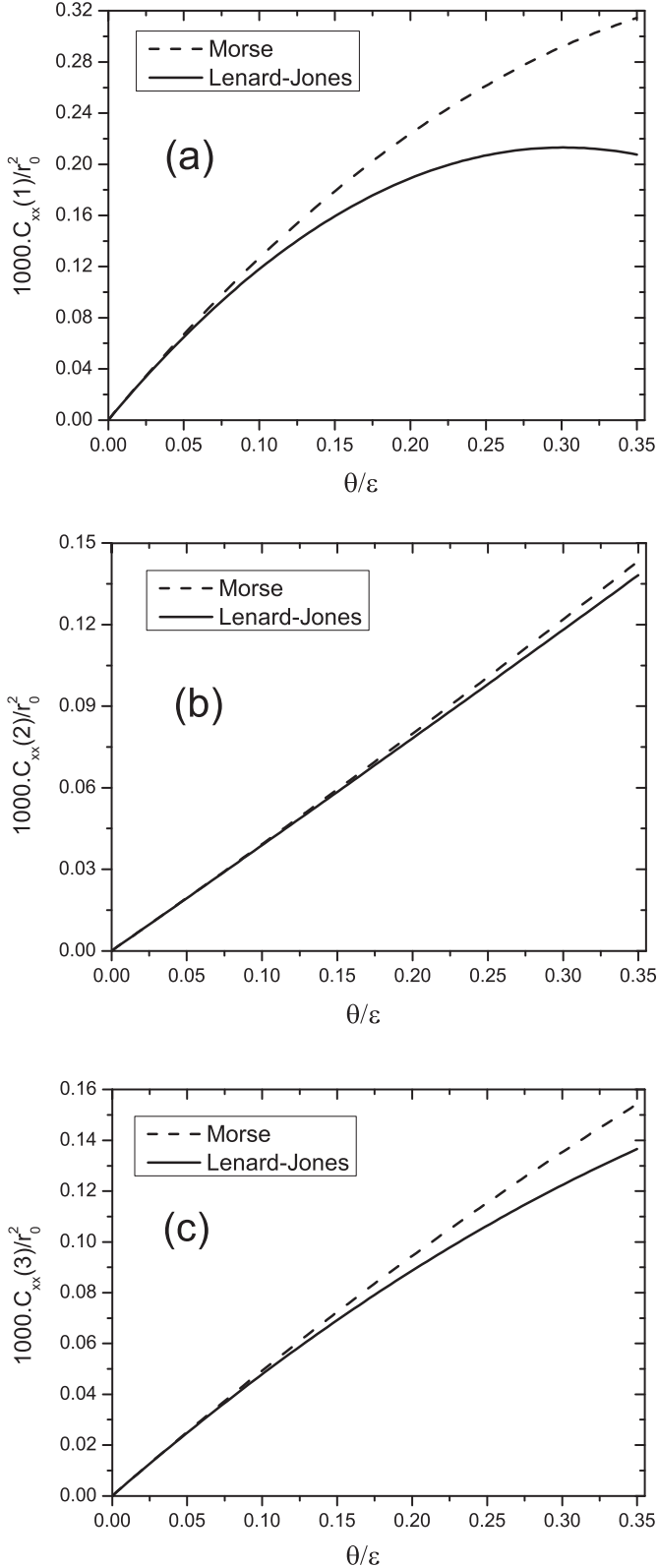

**Fig. 3.** Quadratic correlation moments between atomic displacements in hexagonal lattice for: (a) nearest, (b) second, and (c) third neighbors.

where $\sigma = r_0/2^{1/6}$. The latter potential is more anharmonic and decreases with increasing interatomic distances somewhat slower than former (see Fig. 2).

The values of the derivatives of the interatomic potentials are shown in Table 1.

## 5. Results and discussion

Here, we investigate interatomic correlation moments in two-dimensional model of a weakly anharmonic crystal (i.e., not very high temperatures) with hexagonal lattice, using for the numerical evaluations the Morse and Lenard–Jones potentials. The two-dimensional hexagonal lattice represented in Fig. 1 is the simplest model of a close-packed crystal.

For any pair of atoms, we shall use the coordinate system whose $X$-axis runs through the corresponding lattice points. We calculate the correlation moments of the atomic displacements of the nearest and third neighbors in the coordinate system $X \times Y$, and for second neighbors the system $X' \times Y'$ (see Fig. 1). In such systems: $C_{xy}(1) = 0$, $C_{xy}(2) = 0$ and $C_{xy}(3) = 0$. Basically, the correlations between the longitudinal atomic displacements, in hexagonal lattice, are smaller than those in the linear chain [9,11] and in the square lattice [10]. The interatomic correlation moments decrease as the interatomic distance increases, with it going slower along the line passing through a nearest neighbor of an atom than along other directions. Some results are presented in Fig. 3. Note that the interatomic correlation moments are lower when using Lenard–Jones potential instead of Morse potential, and the anharmonic effects are greater by using Lenard–Jones potential (see Fig. 2).

We intend in the future to apply this study in a graphene monolayer. In this case the most appropriate potential is the "Tewary-potential" [20].